\documentclass{ws-ijmpe}

\begin{document}



\title{MULTIFRAGMENTATION IN COLLITIONS OF \\
4.4GEV-DEUTERONS WITH GOLD TARGET
}

\author {S.P.~AVDEYEV$^\ast$, V.A.~KARNAUKHOV$^\ast$, H.~OESCHLER$^{\ast\ast}$,
V.V.~KIRAKOSYAN$^\ast$, P.A.~RUKOYATKIN$^\ast$,
A.~BUDZANOWSKI$^\parallel$, W.~KARCZ$^\parallel$, E.~NORBECK$^\P$, A.S.~BOTVINA$^\sharp$}

\address{$^\ast$ Joint Institute for Nuclear Research, 141980 Dubna, Russia \\
$^{\ast\ast}$ Institute f\"{u}r  Kernphysik, Darmstadt University of Technology, \\
~~64289 Darmstadt, Germany \\
$^\parallel$ H. Niewodnicza\'{n}ski Institute of Nuclear Physics, 31-342 \\
  ~~Cracow, Poland \\
$^\P$ University of Iowa, Iowa City, IA 52242, USA \\
$^\sharp$ Institute for Nuclear Research, 117312 Moscow, Russia }

\maketitle

\begin{history}
\received{(received date)}
\revised{(revised date)}
\end{history}

\begin{abstract}
The relative velocity correlation function of pairs of intermediate mass fragments has been studied for d+Au collisions at 4.4 GeV. Experimental correlation functions are compared to that obtained by multibody Coulomb trajectory calculations under the assumption of various decay times of the fragmenting system. The combined approach with the empirically modified intranuclear cascade code followed by the   statistical multifragmentation model was used to generate the starting conditions for these calculations. The fragment emission time is found to be less than 40 fm/c.
\\
PACS Number(s): 25.40.-h, 25.70.Mn, 25.70.Pq
\end{abstract}

\section{Introduction}
The main decay mode of very excited nuclei (E* $\geq$ 4 MeV/nucleon) 
is copious emission of intermediate mass fragments (IMF), which are heavier than 
$\alpha$-particles but lighter than fission fragments. An effective way to produce hot nuclei is reactions induced by heavy ions with energies up to hundreds of MeV per nucleon. But in this case the heating of the nuclei may be accompanied by compression, rotation, and shape distortion, which can essentially influence the decay properties of hot nuclei. 
The picture becomes clearer when light relativistic projectiles (protons, antiprotons, pions) are used. In this case, fragments are emitted by only one source (the slowly moving target spectator). Its excitation energy is almost entirely thermal. Light relativistic projectiles provide therefore a unique possibility for investigating thermal multifragmentation.

It has been found experimentally that the process is characterized by two volumes \cite{1,2}.
The first one corresponds to the chemical freeze-out state when prefragments are formed, $V_{t}$ = (2.6 $\pm$ $0.2)V_{0}$. 
The second one is reached by the nucleus after its descent from the top of the fragmentation barrier to the multi-scission point. It is called the kinetic freeze-out volume $V_{f}$ = (5.0 $\pm$ 0.5)$V_{0}$.

The decay properties of hot nuclei are well described by statistical models of multifragmentation\cite{3,4} and this can be considered as an indication that the system is in thermal equilibrium or at least close to that.

The time scale of fragment emission is a key parameter for understanding the decay mechanism of highly excited nuclei. Is it sequential and independent evaporation of IMF's or is it a   multibody decay mode with almost simultaneous emission of fragments governed by the total accessible phase space? As was suggested by D.H.E. Gross in ref.\cite{5} 
``simultaneous''  means that fragments are liberated during a time interval which is smaller than the Coulomb acceleration time $\tau_{c}$, 
when the kinetic energy  of fragments amounts to ~90\% of  the asymptotical  value. 
According to \cite{5}, $\tau_{c}$ is estimated to be (400 - 500) fm/c. 
Fragments emitted within this time interval  are considered being not independent as they interact via the Coulomb force while being accelerated in the electric field of the source. As a result, the yield of events with small relative velocities of the fragments (or small relative angle between them) is suppressed. The magnitude of this effect drastically depends on the emission time since the longer the time separation of the fragments, the larger their space separation and the weaker the mutual Coulomb repulsion. 
Thus, measurement of the IMF emission time $\tau_{em}$ 
(the mean time separation between two fragment emissions in a given event) is a direct way to answer the question as to the nature of the multifragmentation phenomenon. In some papers, the mean lifetime of the fragmenting system $\tau_{s}$ is used to characterize the process of disintegration. There is a simple relation between these two quantities\cite{6,7} 
\begin{equation}
\tau_{em}  = \frac{\tau_{s}}{M-1}\sum_{n=1}^{M-1}\frac{1}{n}. \label{eq:eps}
\end{equation}
Both values are close to each other when the mean IMF multiplicity {\it M} is in the range 2-3, as in the case of the light relativistic projectiles. 

The time scale for IMF emission is estimated by comparing the measured correlation function with the multibody Coulomb trajectory calculations with $\tau_{em}$ as a parameter. There are two procedures to measure the emission time: analysis of the IMF-IMF correlation function of the relative angle or the relative velocity. 

The first measurements of the time scale for the thermal multifragmentation were performed for $^{4}$He+Au collisions at 14.6 GeV by analyzing the IMF-IMF relative angle correlation\cite{6,7}. 
It was found that $\tau_{em}$ is less than 75 fm/c. 
The same procedure was used by the FASA group for the  p+Au interaction at 8.1 GeV\cite{8} when the emission time $\tau_{em} \leq$ 70 fm/c was found. A similar value was obtained by the ISiS collaboration for collisions of 4.8 GeV $^{3}$He with a gold target\cite{9}. In this paper IMF-IMF relative velocity correlations were studied. A general overview of the experimental activity in this field can be found in review paper \cite{10}. 

In the present paper the temporal characteristics of multifragmentation are investigated for the first time for interaction of 4.4 GeV deuterons with the Au target. For that purpose the IMF-IMF correlation function of the relative velocity was studied.


\section{Experimental}

The experiment has been performed with the 4$\pi$ setup FASA\cite{11,12} installed at the external beam of the Dubna superconducting accelerator NUCLOTRON. The FASA device consists of two main parts:
\begin{itemlist}
 \item the array of thirty $\Delta$E-E telescopes, which serve as triggers for the read-out of the whole FASA detector system. These telescopes allow measuring the fragment charge and energy. The spatial distribution of fragments is also obtained (with steps $\Delta\Theta$ = 10$^{\circ}$),
 \item the fragment multiplicity detector (FMD) including 58 thin CsI(Tl) counters (with scintillator thickness around 35mg/cm$^{2}$), which cover 81\% of 4$\pi$. The FMD gives the number of IMF's in the event and their spatial distribution.
\end{itemlist}

The fragment telescopes consist of a compact ionization chamber as the $\Delta$E counter and a Si (Au)   semiconductor detector as the E spectrometer. Effective thickness of the E detector was around 700$\mu$, which is enough to measure the energy spectra of all intermediate mass fragments.  The ionization chambers have a shape of a cylinder (50 mm in diameter, 40 mm in height) and are made from polished brass. The entrance and exit windows are made from organic films ($\sim$100$\mu$g/cm$^{2}$) covered by a thin gold layer prepared by thermal evaporation. A gold wire 0.5 mm in diameter is used as the anode. The cathode (brass cylinder and mechanically supported thin entrance and exit windows) is surely grounded.  Carbon fluoride CF$_{4}$ at the pressure of 50 torr is used as a working gas.

A self-supporting Au target ($\sim$1.5 mg/cm$^{2}$) is located at the center of the FASA vacuum chamber supported by thin tungsten wires. The energy calibration of the counters was done periodically using a precise pulse generator and a $^{241}$Am $\alpha$ source. The beam intensity was around 10$^{10}$ particles per spill. The beam spot was continuously controlled by two multiwire proportional chambers placed at the entrance and the exit of the FASA device. The beam intensity was measured by the special ionization chamber located 150 cm behind the target. The spill length was 1.5 s, the frequency of the beam bursts was 0.1 Hz.

Fig. 1 gives an example of the $\triangle$E-E plot measured by one of the telescopes for d+Au collisions at the beam energy 4.4 GeV. The energy is given for the fragment at the front of the telescope. 
It was obtained from the measured value by adding the energy absorbed on the way from the target 
to the E-detector. The loci for different intermediate mass fragments are well resolved in the range from lithium to silicon. The energy cutoff caused by adsorption of the low energy fragments before the Si(Au) detector increases with Z of fragments. Thus, Ne fragments with the energy lower than 25 MeV do not reach the E-counter.

\begin{figure}[th]
\centerline{\psfig{file=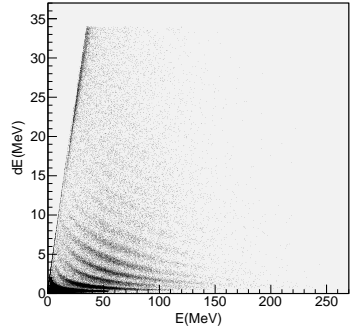,width=5cm}}
\vspace*{8pt}
\caption{$\Delta$E-E plot measured by one of the thirty telescopes of FASA device, 
total number of events detected is around 1.5$\cdot$10$^{6}$.}
\end{figure}

\section{Results of Measurements}

We used a refined version of the intranuclear cascade model (INC)\cite{13} 
to describe the fast stage of the reaction and to get the N, Z and the excitation energy distributions of the target spectators. 
The mean excitation energy of the target spectator is (150-200) MeV according to our analysis\cite{14,15}. 
The excitation energy distribution is rather wide, and 10-15\% of the total inelastic cross-section correspond to the excitation energies of residuals higher than the fragmentation threshold (300-400 MeV). We use the Moscow-Copenhagen Statistical Model (SMM)\cite{3,4} 
to describe the disintegration of hot residuals. The primary fragments are hot, and their de-excitation is also considered by the 
SMM to get the final distributions of cold IMF's. The last stage of the reaction is Coulomb expansion of the final system, 
when fragments are getting their kinetic energy. 
This process is described by the multibody trajectory calculations for fragments in the freeze-out volume $V_{f}$ = 3$V_{0}$. 
This part of the combined model gives the IMF kinetic energy spectra and velocity distributions for fragments.

It was shown in our papers\cite{14,15} 
that traditional (INC + SMM) model calculations failed to describe the data for the IMF multiplicities. 
It was considered as anindication that the cascade calculations overestimate the excitation energies of the residues. In order to overcome this difficulty the calculated excitation energies are reduced empirically by the factor $\alpha$ on the event-by-event basis

\begin{equation}
\alpha  = \frac{<M_{exp}>}{<M_{INC+SMM}>} \label{eq:eps}
\end{equation}
where $<M_{exp}>$ - measured mean multiplicities for events with at least one IMF.
The model calculated mean IMF multiplicity is given as a denominator. 
The mass loss during the intra-nuclear cascade is reduced by the factor\cite{15} 
(1 - $\alpha$). As a result, one gets a INC+Exp+SMM model which we 
used in the present paper.

\begin{figure}[th2]
\centerline{\psfig{file=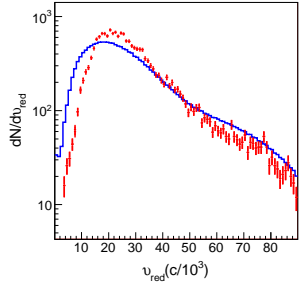,width=5cm}}
\vspace*{8pt}
\caption{The yield as a function of the reduced relative velocity of coincident IMF's (points). Histogram is for the mixed events.}
\end{figure}

Figure 2 shows the distribution of the reduced relative velocity of the fragments. 
\begin{equation}
v_{red} = v_{rel}/\sqrt{Z_{1}+Z_{2}} \label{eq:eps}
\end{equation}
Difference of velocities of two fragments from the same event is normalized by the square root of the sum the fragment charges. This normalization was suggested in \cite{16,17,18} 
to compensate charge dependence of the velocity difference 
$\mid{\it v}_i$-${\it v}_k\mid$. 
It is measured for 18129 events with detection of at least two coincident fragments (the total number of triggers was 4$\cdot10^6$). 
Data for all the fragments with Z between 3 and 20 are included.

The yield of a pairs with small ${\it v}_{red}$ 
is remarkably suppressed. It is partly related to the phase space factor (or phase space density), as is demonstrated in fig.2 by the histogram obtained for the 
``mixed events'' (i.e. for a pair of fragments from different collisions). But the Coulomb repulsion between the coincident fragments makes suppression at ${\it v}_{red}\approx$ 0 much more significant. 

Traditionally, Coulomb interaction of genetically related fragments is described in the terms of the correlation function 
\begin{equation}
1 + R(v_{red})  = CN_{corr}(v_{red})/N_{uncorr}(v_{red}), \label{eq:eps}
\end{equation}
where $N_{corr}(v_{red}$) is the measured coincidence yield. 
The denominator is composed of the mixed events, C is the normalization constant.

\section{Discussion}

The model correlation functions are obtained after multibody trajectory calculations for the fragments in the Coulomb field of the system. It is made on an event by event basis using the INC+Exp+SMM model for describing specification of events (Z, A and excitation energy of IMF source). 

\begin{figure}[th3]
\centerline{\psfig{file=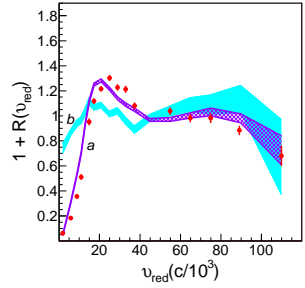,width=5cm}}
\vspace*{8pt}
\caption{Reduced relative velocity correlation function of the coincident fragments pairs. The points are experimental data, lines are obtained with INC+Exp+SMM combined model followed by the multi-body Coulomb trajectory calculations with decay time of fragmenting system: a - prompt, b - 160 fm/c.}
\end{figure}

The starting time for each fragment to move along a Coulomb trajectory was randomly chosen according to the decay probability of the system, P(t)$\sim$exp(-t/$\tau_s$). 
The calculations were performed for the emission times between 0 and 200 fm/c 
and for freeze-out volume $V_{f}$ = 3$V_{0}$, where $V_{0}$ is the nuclear volume at normal density. The experimental filter was applied to be in line with the experimental definition of the correlation function. It is seen from Fig. 3 that the model calculations reproduce well the deep drop of the yield for small relative velocities. 

For the quantitative comparison of the data with the model calculations the Pirson criterion is used. The plot of $\chi^2$ and confidence level CL versus the mean 
\begin{figure}[th4]
\centerline{\psfig{file=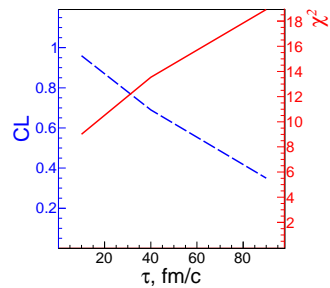,width=5cm}}
\vspace*{8pt}
\caption{Comparison result of the measured and model calculated correlation functions versus emission time. 
CL - dashed line, $\chi^2$ - solid line.}
\end{figure}
decay time of the system is shown in Fig. 4. The mean life time of the fragmenting system is equal to or less than 40 fm/c at the confidence level $\sim$67\%. This result is in agreement with the emission time values obtained in the previous studies of the thermal multifragmentation\cite{6,7,8,9}. 

\section*{Acknowledgements}

The authors thank A.I. Malakhov, A.G. Olshevski, I.N. Mishustin and W. Trautmann for the illuminating discussions. The research was supported in part by the Russian Foundation for Basic Research, Grant No. 06-02-16068, by the Grant of the Polish Plenipotentiary to JINR, by Bundesministerium f\"{u}r Forschung und Technologie, Contract No. 06DA453.

\end{document}